\documentstyle[aps,preprint,epsfig,floats,eqsecnum]{revtex}
\def\laq{\ \raise 0.4ex\hbox{$<$}\kern -1.1em\lower 0.62
ex\hbox{ $\sim$}\ }
\def\gaq{\ \raise 0.4ex\hbox{$>$}\kern -0.7em\lower 0.62
ex\hbox{$\sim$}\ }
\def\dslash{\ \lower 0.4ex\hbox{$\nabla$}\kern -1.0em\lower 0.2
ex\hbox{ $\slash$}\ }
\def\N{{\cal N}}

\begin{document}

\title{Causal Entropy Bound for Non-Singular Cosmologies} \author{Ram
  Brustein${}^{(1)}$ Stefano Foffa${}^{(1)}$ and Avraham E.
  Mayo${}^{(2)}$} \vskip 2 cm \address{(1) Department of Physics,
  Ben-Gurion University,\\ Beer-Sheva 84105, Israel \\ (2) The Racah
  Institute of Physics,Hebrew University of Jerusalem,\\ Givat Ram,
  Jerusalem 91904, Israel \\ {\rm E-mail:} {\tt
    ramyb@bgumail.bgu.ac.il, foffa@bgumail.bgu.ac.il,
    mayo@cc.huji.ac.il} }

\maketitle

\begin{abstract}
The conditions for validity of the Causal Entropy Bound (CEB) are
verified in the context of non-singular cosmologies with
classical sources. It is shown that they are the same conditions
that were previously found to guarantee validity of the CEB: the
energy density of each dynamical component of the cosmic fluid
needs to be sub-Planckian and not too negative, and its equation
of state needs to be causal. In the examples we consider, the CEB
is able to discriminate cosmologies which suffer from potential
physical problems more reliably than the energy conditions
appearing in singularity theorems.

\end{abstract}

\section{introduction}

The validity of entropy bounds in Bekenstein's non-singular
cosmological model \cite{bmodel} has been recently challenged
\cite{mayo}. In the course of the analysis some of the energy
conditions that sources in the Einstein equations are assumed to
obey \cite{Bousso,Wald} were questioned. Here we determine the
conditions that guarantee the validity of the CEB \cite{CEB} for
non-singular cosmologies with classical sources, and discuss their
relation to the energy conditions of the classic singularity
theorems\cite{singth}.

CEB is a covariant entropy bound which is applicable, in principle, to
any space-like region \cite {CEB} in an arbitrary space-time dimension
$D$ \cite{BFV}. It is an improvement of the Hubble Entropy Bound (HEB)
\cite{GV1} (see also \cite{EL,BR,KL}), which was motivated by the
following reasonable assumptions $(i)$ entropy is maximized by the
largest stable black hole that can fit in a given region of space.
$(ii)$ the largest stable black hole in a cosmological background is
typically of size comparable to that of the Hubble horizon (this
assumption is qualitatively supported by previous calculations
\cite{Carr}). In cosmological backgrounds, the CEB refines HEB by
defining the ``horizon" concept through the identification of a
critical (``Jeans"-like) causal connection scale $R_{\rm CC}$, above
which perturbations are causally disconnected, so that black holes of
larger size are unlikely to form.

In homogeneous and isotropic $D$ dimensional cosmological backgrounds
$R_{\rm CC}$ depends on the Hubble parameter $H(t)$, its
time-derivative $\dot{H}(t)$, and the scale factor $a(t)$ \cite{BFV},
\begin{eqnarray}
\label{rcch} R_{\rm CC}^{-2}&=&\frac{D-2}{2}{\rm Max}\left[
\dot{H} + \frac{D}{2}H^2 + \frac{D-2}{2}\frac{k}{a^2}, -\dot{H} +
\frac{D-4}{2}H^2 + \frac{D-2}{2}\frac{k}{a^2}\right] \nonumber \\
&=&\frac{4\pi G_{\rm N}}{D-1} {\rm Max} \biggl[
 \rho - (D-1)p\, , (2D-5)\rho + (D-1)p
 \biggr],
\end{eqnarray}
where $k=0,\pm1$ determines the spatial curvature. To derive the
second equality we have used Einstein's equations, $G_{\mu\nu} = 8 \pi
G_N T_{\mu\nu}$ and a perfect-fluid form for the energy-momentum
tensor. Notice that $R_{\rm CC}$ is well defined if $\rho$ is positive
because the maximum in Eq.(\ref{rcch}) is larger than the average of
the two entries in the brackets, and the average is equal to
$2(D-2)\rho$.

Previously \cite{CEB} three cases which were believed to exhaust
all possible types of cosmologies were considered \footnote{In
\cite{CEB} space curvature was implicitly included in the total
energy budget as a regular additional source.}:
\begin{enumerate}
\item $|\dot H|\sim H^2 \sim |k|/a^2$, or $|\dot H|\sim H^2 \gg |k|/a^2$.
  In this case effective energy density and
  pressure are of the same order, $\rho\sim p$, and all length scales
  that may be considered in entropy bounds, such as particle horizon,
  apparent horizon, $R_{\rm CC}$, and the Hubble length, are
  parametrically equal. This case includes non-inflationary FRW
  universes with matter and radiation.
\item $H^2 \gg |k|/a^2,|\dot H|$. In this case $|\rho+ p| \ll \rho$,
and the
  universe is inflationary.  Here the naive holographic bound fails
  miserably, but HEB, CEB and Bousso's modification of the
  holographic entropy bound \cite{Bousso} do well.
\item $|\dot H|\gg H^2,\ |k|/a^2$. In this case \ $|\rho| \ll p$.  Since $\rho$
  and $p$ are the effective energy density and pressure, there are no
  problems with causality. This case occurs, for instance, near the
  turning point of an expanding universe which recollapses, or
  near a bounce of a contracting universe which reexpands.

  \hspace{-.5in} \hbox{There is however an additional case which was
    not explicitly included in cases (1)-(3):}

\item $k/a^2 \gg |\dot H|,\ H^2$ so that spatial curvature
  determines the causal connection scale. This occurs, for example,
  when both $H$ and $\dot H$ vanish as in a closed Einstein Universe,
  or in the static version of Bekenstein's non-singular Universe
  \cite{bmodel}.
\end{enumerate}

\noindent Here we discuss this last case and show that the same
conditions that guarantee validity of CEB in the first three
cases suffice to guarantee its validity in the fourth case.

CEB states that the maximal entropy $S_{\rm CEB}$ that can be
contained in a space-like region of proper volume $V$ is given by
(our units are such that $\hbar=c=1$ and $G_{\rm N}=M_{\rm
P}^{-(D-2)}= \ell_{\rm
  P}^{D-2})$,
\begin{eqnarray}
\label{scebD} S_{\rm CEB}=\beta n_{\rm H} S^{\rm BH}=\beta
\frac{V}{V(R_{\rm CC})} \frac{{\cal A}(R_{\rm CC})}{4 \ell_{\rm
P}^{D-2}},
\end{eqnarray}
where $n_{\rm H} \equiv \frac{V}{V(R_{\rm CC})}$ is the number of
causally connected regions in the volume considered, $V(x)$ denotes
the volume of a region of size $x$, ${\cal A}(x)$ denotes the area of
this region, and $\beta$ is a fudge factor reflecting current
uncertainty on the actual limiting size for black-hole stability. For
a spherical volume in flat space we have $V(x)=\Omega_{D-2}
x^{D-1}/(D-1)$, and ${\cal A}(x)=\Omega_{D-2} x^{D-2}$, with
$\Omega_{D-2}=2\pi^{(D-1)/2}/ \Gamma\left(\frac{D-1}{2}\right)$, but
in general the result is different and depends on the
spatial-curvature radius. Since $\frac{{\cal A}(R_{\rm CC})}{V(R_{\rm
    CC})}\sim \frac{D-1}{R_{\rm CC}}$,
\begin{equation}
\label{scebE}
 S_{\rm CEB}=\alpha (D-1) \frac{V}{G_N R_{\rm CC}},
\end{equation}
where $\alpha$ is a numerical parameter of order one.

Conditions for validity of CEB were determined in \cite{CEB,BFV}.
Loosely speaking, energy densities are required to be sub-Planckian,
and the total energy density of the cosmic fluid is required to be
positive. In particular, for a universe with a large number of fields
$\N$, in thermal equilibrium at temperature $T$, the CEB was found to
be valid for temperatures not exceeding a value of order $M_{\rm
  P}/\N^{{1\over D-2}}$ (see also \cite{Bek3,GSL}).

\section{CEB in non-singular cosmologies}

\subsection{Einstein Universe with radiation}

The simplest example of a non-singular cosmology is a static
Einstein model in $D$ dimensions. This model requires positive
curvature, and two types of sources: cosmological constant and
dust; we denote by $\rho_{\Lambda}$ and $\rho_{\rm m}$ the energy
densities associated with each of the two components. To provide
entropy we need an additional source, which we choose to be
radiation consisting of $\N$ species in thermal equilibrium at
temperature $T$. The energy density of the radiation is given by
$\rho_{\rm r}=\N T^D$, and the entropy density of the radiation
is given by $s_{\rm r}=\N T^{D-1}$ (we ignore here numerical
factors since we will be interested in scaling of quantities).
The total entropy of the system is given entirely by the entropy
of the radiation $S_{\rm r}=s_{\rm r} V$.

In term of these sources, Einstein's equations can be written in the
following way:
\begin{eqnarray}\label{ein1}
H^2 + \frac{1}{a^2}&=& \frac{16 \pi G_{\rm N}}{(D-2)(D-1)}\rho_{\rm tot}=
\frac{16 \pi G_{\rm N}}{(D-2)(D-1)}\left(\rho_{\Lambda}+\rho_{\rm m}+
\rho_{\rm r}\right)\\
\label{ein2}
\dot{H} - \frac{1}{a^2}&=&
-\frac{8\pi G_{\rm N}}{(D-2)}\left(\rho_{\rm tot} + p_{\rm tot}\right)=
-\frac{8\pi G_{\rm N}}{(D-2)(D-1)}
\left[D\rho_{\rm r}+(D-1)\rho_{\rm m}\right]\, ,
\end{eqnarray}
where we have used in Eq.~(\ref{ein2}) the equations of state
relating pressure to energy density:
$p_{\Lambda}=-\rho_{\Lambda}$, $p_{\rm m}=0$, and $(D-1)p_{\rm
r}=\rho_{\rm r}$.

For given $\rho_{\rm m}$ and $\rho_{\rm r}$, one can choose
$\rho_{\Lambda}$ and the scale factor $a$ such that $H$ and $\dot{H}$
vanish in Eqs.~(\ref{ein1}) and (\ref{ein2}), and thus obtains a
static solution. In particular, the condition given by
Eq.~(\ref{ein2}) determines the scale factor in terms of $\rho_{\rm
  m}$ and $\rho_{\rm r}$,
\begin{equation}\label{e2vanish}
a^2=\frac{(D-2)(D-1)}{8\pi G_{\rm N}}\frac{1}{D\rho_{\rm r} + (D-1)\rho_{\rm m}}\, .
\end{equation}
Note that since both $H$ and $\dot H$ vanish identically, $R_{\rm CC}$
is determined solely by the scale factor $a$ given in
Eq.(\ref{e2vanish}), as discussed previously.

We now wish to determine under which conditions (if any) some
violations of CEB may occur in this model. Recall that according
to Eq.(\ref{scebE}) the CEB bounds the total entropy of a region
contained in a comoving volume $V$ by $S_{\rm CEB}=\alpha
(D-1)\frac{V}{G_{\rm N} R_{\rm CC}}$,  and that in the static case
under consideration $R_{\rm CC}=2 a/(D-2)$. The square of the
ratio of $S_{\rm CEB}$ and the entropy of the system $S_{\rm r}$,
is given by
 \begin{eqnarray}
 \label{Soverceb}
\left(\frac{S_{\rm CEB}}{S_{\rm r}}\right)^2 &=& \left(
\frac{\alpha(D-1)} {s_{\rm r} R_{\rm CC}
G_{\rm N}}\right)^2 \nonumber \\
&=& \left[ 2\pi \alpha^2 (D-1)(D-2)\right] \left[{D +
(D-1){\rho_{\rm m}\over \rho_{\rm r}}}\right] \left[\frac{1}{\N}
\left(\frac{M_P}{T}\right)^{D-2}\right] \, .
\end{eqnarray}
Since the second factor in expression (\ref{Soverceb}) is larger than
unity if $\rho_{\rm m}$ and $\rho_{\rm r}$ are positive, and
neglecting the overall prefactor which is independent of the sources
in the model, we conclude that the CEB is valid provided that
\begin{equation}\label{subP}
 \N\left({T\over M_{\rm P}}\right)^{D-2}\laq 1.
\end{equation}
This is the same condition discussed in \cite{BFV}, and should be
interpreted as a requirement that temperatures are sub-Planckian, in
the case of many number of species $\N$ (see also \cite{Bek3,GSL}).

We therefore conclude that, as long as the temperature of
radiation stays well below  Planckian, CEB is upheld. The fact
that the model is gravitationally unstable to matter
perturbations does not seem to be particularly relevant to the
issue of validity of the CEB.

\subsection{Bekenstein's Universe}
A non-singular cosmological model which can describe
time-dependent cosmologies was found years ago by
Bekenstein\cite{bmodel}. This is a 4D Friedman-Robertson-Walker
universe which is conformal to the closed Einstein Universe. It
contains dust, consisting of $N$ particles of mass $\mu$ ($N$ is
constant and $\mu$ is positive), coupled to a classical conformal
massless scalar field $\psi$, and $\N$ species of radiation in
thermal equilibrium. The action for the dust-$\psi$ system is
given by
\begin{eqnarray}
{\cal S}=-\frac{1}{2}\int\sqrt{-g}\left[\left(\nabla\psi\right)^2 +
\frac{1}{6} \psi^2 R\right]d^4 x - \int\left(\mu +
f\psi\right)d\tau.
\end{eqnarray}
It includes in addition to the usual action for free point
particles of rest mass $\mu$, a dust-scalar field interaction
whose strength is determined by the coupling $f$. Accordingly, we
may define the effective mass of the dust particles: $\mu_{\rm
eff}=\mu+f \psi$.

The total energy density and pressure in Bekenstein's Universe
are given by
\begin{eqnarray}\label{rototmay}
\rho_{\rm tot}=\rho_{\rm r}+\rho_{\psi}+\rho_{\rm m},\quad
p_{\rm tot}&=&p_{\rm r}+p_\psi+p_{\rm m} ,
\end{eqnarray}
where $\{\rho_{\rm r},p_{\rm r}\}$, $\{\rho_{\psi},p_\psi\}$, and
$\{\rho_{\rm m},p_{\rm m}\}$ are the energy densities and
pressures associated with the radiation, scalar field and dust
respectively. They depend on the scale factor in the following way
\begin{eqnarray}
\label{equations}
\rho_{\rm r}&=&{\cal C} \N a^{-4}=\N T^4, \nonumber \\
 \rho_\psi &=& {1\over 2}f^2N^2 a^{-4},  \\
 \rho_{\rm m} &=&N\mu_{\rm eff}
a^{-3}=N\mu a^{-3} -2\rho_\psi, \nonumber
\end{eqnarray}
and their equations of state $\gamma_{\rm r}=p_{\rm r}/\rho_{\rm
r}$, $\gamma_{\psi}=p_\psi/\rho_{\psi}$,  $\gamma_{\rm m}=p_{\rm
m}/\rho_{\rm m}$ are the following
\begin{eqnarray}
\label{eoss}
\gamma_{\rm r}&=&1/3, \nonumber \\
 \gamma_\psi &=& -1/3,  \\
 \gamma_{\rm m} &=&0. \nonumber
\end{eqnarray}
The dependence of $\psi$ on $a$  $\psi=-f N a^{-1}$,
yields $\mu_{\rm eff}=\mu-f^2 N a^{-1}$. ${\cal C}$ is an
integration constant and the only source of entropy is the
radiation whose entropy density is given by $s_{\rm r}=\N T^3$.

The solution for the scale factor $a$ is given in terms of the
conformal time $\eta$ by
\begin{equation}
a(\eta)=a_0(1+B \sin\eta).
\end{equation}
We assume that $a_0$, the mean value of the scale factor, is
macroscopic, so it is large in our Planck units. If $B=0$ the
solution describes a static universe very similar to the closed
Einstein Universe discussed previously. For $0<B<1$ the solution
describes a ``bouncing universe": the universe bounces off at
$\eta=3 \pi/2$ when the scale factor is minimal $a=a_{\rm
min}=a_0(1-B)$, expands until it turns over at $\eta=5\pi/2$ when
its scale factor is maximal $a=a_{\rm max}=a_0(1+B)$, and
continues to oscillate without ever reaching a singularity. The
equations of motion require that the energy densities of the
sources obey the following equalities at all times \cite{bmodel}:
\begin{equation}
2 {a\over a_0} \left({\rho_\psi-\rho_{\rm r}\over2\rho_\psi+
\rho_{\rm m}} \right)=1-B^2={a_{\rm min} a_{\rm max}\over
{a_0}^2}.
 \label{equality}
\end{equation}
Since $2\rho_\psi+\rho_{\rm m}=N\mu a^{-3}>0$, $\rho_{\rm r}>0$,
and $B^2<1$, it follows that a necessary condition for a bounce is
that $\rho_{\rm r}<\rho_\psi $. This implies that the total
pressure $\frac{1}{3} (\rho_{\rm r}-\rho_{\psi})$ is always
negative. Moreover, Eq.(\ref{equality}) for $a=a_{\rm min}$
implies that $\rho_{\rm m}\leq-2\rho_{\rm r}<0$ there. But then,
the conclusion must be that in order to avoid a singularity,
$\mu_{\rm eff}<0$ at least at the bounce. It is possible,
however, to find a range of initial conditions and parameters such
that $\mu_{\rm eff}$ is positive near the turnover.

The result that $\rho_{\rm r}$ and $\rho_\psi$ are manifestly
positive definite, but $\rho_{\rm m}$ can (and in fact must) be
negative some of the time, suggest that it might be possible to
parametrically decrease $\rho_{\rm tot}$ by lowering $\mu_{\rm
eff}$ (making it large and negative) by
increasing the coupling strength $f$, so that the amounts of
radiation and entropy are kept constant. As it turns out this is
exactly the case in which the CEB can be potentially violated.
Using Einstein's equations to express $R_{\rm CC}$ in terms of
the total energy density and pressure, we find the ratio
$\left(S_{\rm CEB}/S_{\rm r}\right)^2$:
\begin{eqnarray}
 \label{rat1}
 \left({S_{\rm CEB}\over S_{\rm r}}\right)^2 \sim {G_{\rm N}}^{-2}
 \left(\frac{\rho_{\rm r}}{\cal N}\right)^{-3/2}
 \frac{1}{{\cal N}^2} G_{\rm N}\,
 {\rm Max} \left[
 {\rho_{\rm tot}\over 3}-p_{\rm tot} ,\rho_{\rm tot}+p_{\rm tot}
 \right].
\end{eqnarray}
A system for which the ratio above is smaller than one would violate
the CEB.  Recalling that the maximum on the r.h.s. of (\ref{rat1}) is
always larger than the mean of the two entries and rearranging we find
\begin{equation}
\left({S_{\rm CEB}\over S_{\rm r}}\right)^2 \gaq
\left[\frac{1}{\N}\frac{M_P^2}{T^2}\right]{\rho_{\rm tot}\over
\rho_{\rm r}}. \label{ratio}
\end{equation}
Since we assume that the model is sub-Planckian, namely that the first
factor is larger than one as in Eq.(\ref{subP}), the only way in which
CEB could be violated is if somehow the second factor was
parametrically small. As discussed above, it does seem that the second
term $\rho_{\rm tot}/\rho_{\rm r}$ can be made arbitrarily small by
decreasing $\rho_{\rm tot}$ while keeping $\rho_{\rm r}$ constant.
Consequently, it is apparently possible to make the ratio $S_{\rm
  CEB}/ S_{\rm r}$ smaller than one and obtain a CEB violating
cosmology. But this can be achieved only if the effective mass of the
dust particles is negative (and large) as can be seen from
Eq.~(\ref{rototmay}).

Violations of the CEB (and as a matter of fact, of any other
entropy bound such as Bekenstein's \cite{BEB}, or Bousso's
\cite{Bousso}) go hand in hand with large negative energy
densities in the dust sector. In the model under discussion, this
manifests itself in the form of dust particles with highly
negative effective masses. Occurrence of such negative energy
density would most probably render the model unstable (see
below).  We argue that any analysis of entropy bounds should be
performed for stable models. This is particularly relevant for the
CEB, whose definition involves explicitly the largest scale at
which stable black holes could be formed. Note, however, that
instability does not necessarily lead to violations of the CEB as
in the previous case.

To support this argument let us outline possible instabilities in the
dust scalar field system when the dust particles mass is negative. To
do this we need to be more specific about the model.  Consider a
possible field theoretic model for the dust as a fermionic field
$\chi$. In this case the dust-scalar field action is given by
\begin{eqnarray}
{\cal S}=-\frac{1}{2}\int d^4 x \sqrt{-g}\left[\left(\nabla\psi\right)^2
+ \frac{1}{6} \psi^2 R + i \bar\chi \dslash \chi + \mu \bar\chi
\chi + f \psi \bar\chi\chi\right].
\end{eqnarray}
The equations of motion determine the constant non-vanishing
values of $\psi$ (for simplicity consider the static case only)
and $\bar\chi \chi$. We see that when the effective mass $\mu+ f
\psi$ becomes negative the model becomes unstable due to $\chi$
pair production, and will prefer a state with a $\bar\chi\chi$
condensate, which will feed back into $\psi$. Correspondingly,
such rapid creation of pairs would be accompanied by strong
fluctuations. It is not clear whether under these circumstances
the condition for bounce $\rho_{\rm r}<\rho_\psi$ will continue
to hold indefinitely, or whether a collapse to a singularity will
ensue after a finite number of cycles of the universe.  A complete
discussion of the time-dependent situation is beyond the scope of
this paper but it is clear that violations of CEB are related
with a potential instability in the dust sector, and cannot be
simply taken as a bona-fide example of CEB violation.

The fact that Bekenstein's universe is non-singular indicates
that the singularity theorems of Penrose and Hawking
\cite{singth} are somehow eluded.  And indeed the Strong Energy
Condition (SEC) is violated in the model: $\rho_{\rm tot}+3p_{\rm
tot}=2\rho_{\rm r}+\rho_{\rm m}$ is negative at the bounce,
positive at the turnover and changes continuously in between. As
we show later, violation of some energy conditions does not
necessarily mandate a violation of the CEB. We will argue that in
this sense the CEB has a better discriminating power than energy
conditions (see below).

\section{Conditions for validity of CEB \\ with general classical sources}

We may summarize the lessons of the previous examples by imposing
conditions on sources in a generic cosmological setting such that
CEB is obeyed. This analysis is not restricted to a static
universe, nor to a closed one, and contains the previous examples
as particular cases.

We consider a cosmic fluid consisting of radiation, an optional
cosmological constant, and additional unspecified classical
dynamical sources which do not include any contributions from the
cosmological constant or radiation. For simplicity we assume that
the additional sources have negligible entropy. This is the most
conservative assumption: if some of the additional sources have
substantial entropy our conclusions can be strengthened. We use
the previous notations for the total, cosmological, and radiation
energy densities, $\rho_{\rm tot}$, $\rho_{\Lambda}$ and
$\rho_{\rm r}$ respectively, and denote by $\rho^*$ the combined
energy density of the additional sources. Thus
\begin{equation}
\rho_{\rm tot}=\rho_{\rm r}+\rho_{\Lambda}+\rho^*.
\end{equation}
We use the same notation for the relative pressures, and for the
equation of state $\gamma^*\equiv\rho^*/p^*$, which may be
time-dependent.

In term of these sources, the causal connection scale can be written
as
\begin{eqnarray}\label{Rccrho}
&& R_{\rm CC}^{-2}=\frac{4\pi G_{\rm N}}{D-1}\times\nonumber\\
&&{\rm Max} \Biggl\{D\rho_{\Lambda} + \biggl[1 -
(D-1)\gamma^*\biggr]\rho^*\, , (D-4)\rho_{\Lambda} + \biggl[(2D-5) +
(D-1)\gamma^*\biggr]\rho^* + 2(D-2)\rho_{\rm r}\Biggr\}.
\end{eqnarray}

We may now express the ratio of $(S_{\rm CEB}/S_{\rm r})^2$,
neglecting as usual prefactors of order one
\begin{eqnarray}
&& \left(\frac{S_{\rm CEB}}{S_{\rm r}}\right)^2 \sim
\frac{1}{\N}\left(\frac{M_{\rm P}}{T}\right)^{D-2} \times
\nonumber \\ && {\rm Max}\left\{ D\frac{\rho_{\Lambda}}{\rho_{\rm r}}+
\biggl[ 1 - (D-1)\gamma^*\biggr]\frac{\rho^*}{\rho_{\rm r}}\,,
(D-4)\frac{\rho_{\Lambda}}{\rho_{\rm r}}+\biggl[ (2D-5) +
(D-1)\gamma^*\biggr]\frac{\rho^*}{\rho_{\rm r}} +
2(D-2)\right\} .\label{ratiogen}
\end{eqnarray}
As was already pointed out in the previous section, a 
condition for any CEB violations is that this ratio be
parametrically smaller than one. Notice that the first factor is
larger than one by our requirement that the radiation energy
density be sub-Planckian. Thus the only remaining possibility for
violating CEB is that the second factor be parametrically smaller
than unity. As we show below, this can occur only if at least one
of the additional sources has negative energy density.

The r.h.s. of (\ref{ratiogen}) is larger than the average of the two
entries, so that
\begin{equation}
\label{newrat} \left(\frac{S_{\rm CEB}}{S_{\rm r}}\right)^2 \gaq
{1\over\N}\left({M_{\rm P}\over T}\right)^{D-2} (D-2) {\rho_{\rm
tot}\over\rho_{\rm r}}
\end{equation}
Therefore, since $\rho_{\rm tot}>0$, a necessary condition for
this expression to be smaller than unity is that $\rho_{\rm
  tot}\ll\rho_{\rm r}$, which we may reexpress as
\begin{eqnarray}\label{condL}
\frac{\rho_{\Lambda}}{\rho_{\rm r}}\sim
-\left(1+\frac{\rho^*}{\rho_{\rm r}}\right).
\end{eqnarray}
This is not a sufficient condition since the equations of motion
could dictate, for example, that the first factor on the r.h.s.
of eq.(\ref{newrat}) could be parametrically larger than unity at
the same time. By substituting condition (\ref{condL}) into
Eq.~(\ref{ratiogen}), we obtain
\begin{eqnarray}
&& \left(\frac{S_{\rm CEB}}{S_{\rm r}}\right)^2 \sim
\frac{1}{\N}\left(\frac{M_{\rm P}}{T}\right)^{D-2} \times
\nonumber \\ && {\rm Max}\left\{-\biggl[(D-1)(1+\gamma^*)\frac{\rho^*}
{\rho_{\rm r}} + D\biggr]\,,(D-1)(1+\gamma^*)\frac{\rho^*}
{\rho_{\rm r}} + D\right\} .\label{ratnongen}
\end{eqnarray}
Therefore, an additional necessary  condition for $S_{CEB}/S_{\rm
r}$ to be smaller than one is that
\begin{eqnarray}\label{cond*}
(1 + \gamma^*)\rho^*\simeq-\frac{D}{(D-1)}\rho_{\rm r}\, .
\end{eqnarray}
Condition (\ref{cond*}) can be satisfied in two ways:

(i) $1+\gamma^*>0$ and $\rho^*<0$. This obviously requires that at
least one of the sources has negative energy density. In this
case (barring pathologies) the magnitude of $\rho^*$ is
comparable to that of $\rho_{\rm r}$.

(ii) $1+\gamma^*<0$ and $\rho^*>0$. However, for classical
dynamical sources,  this typically clashes with causality which
requires that the pressure and energy density of each of the
additional dynamical sources obey $|p_i|<|\rho_i|$; hence if all
$\rho_i>0$ then necessarily \hbox{$\gamma^*=\left(\sum
p_i\right)/\left(\sum \rho_i\right)>-1$}.

Consequently, condition (\ref{cond*}) cannot be satisfied if all
of the dynamical sources have positive energy densities and
equations of state $|\gamma_i|\le 1$. Bekenstein's Universe
discussed in the previous section fits well within our framework:
the total energy density is positive, but the overall
contribution to $\rho_{\rm tot}$ of all the sources, excluding
radiation (since the cosmological constant vanishes in this
case), is negative and almost cancels the contribution of
radiation, leaving a small positive $\rho_{\rm tot}$.

To summarize, if all dynamical sources (different from the
cosmological constant) have positive energy densities $\rho_i >
0$ and have causal equations of state ($|\gamma_i|\le 1$), and if
radiation temperatures are sub-Planckian, CEB is upheld.

The CEB (and entropy bounds in general) refines the classic
singularity theorems in that it allows cosmologies for which the
singularity theorems are not applicable because some of the
energy conditions are violated, but do not seem to be problematic
in any of their properties, or indicates possible problems already when
the singularity theorems seem perfectly valid. For example, the
scale factor for a closed deSitter Universe (i.e. a closed
Universe containing a positive cosmological constant $\Lambda$)
in $D=4$ is given by
$a(t)=(\frac{\Lambda}{3})^{-1/2}\cosh{\sqrt{\frac{\Lambda}{3}}t}$,
showing a bounce at $t=0$. This is not surprising since the
sources of this model violate the SEC. The reliability of the SEC
as a criterion of discriminating physical and unphysical
solutions is therefore questionable (as is well known in the
context of inflationary cosmology). Alternatively, in a
contracting 4D radiation dominated universe, the singularity
theorems imply the the solution will reach a future singularity,
but the CEB indicates problems already when $T\sim M_{\rm P}/\N^{1/2}$.

In general, the total energy-momentum tensor of a closed
``bouncing" universe violates the SEC, but it can obey the CEB. In
order to see this explicitly  let us consider the ``bounce''
condition, i.e.~ $H=0$, $\dot{H}>0$ for a closed Universe; by
using the Einstein equations (\ref{ein1}-\ref{ein2}), we can
express this condition in terms of the sources as follows:
\begin{eqnarray}\label{bounce}
\rho_{\rm tot}>0,\quad (D-3)\rho_{\rm tot}+(D-1)p_{\rm tot}<0.
\end{eqnarray}
The second of these conditions is (in $D=4$) precisely the
condition for violation of the SEC.  In terms of $\rho_{\rm r}$,
$\rho_{\Lambda}$ and $\rho^*$ this reads
\begin{equation}
\label{condbounce}
2 \rho_{\Lambda} - (D-2)\rho_{\rm r} - \biggl[(D-3) +
(D-1)\gamma^*\biggr]\rho^*>0\, .
\end{equation}
In comparison, a necessary condition that the CEB is violated can
be obtained from Eqs.(\ref{condL}) and (\ref{cond*}),
\begin{equation}
\label{3.10}
2 \rho_{\Lambda} - (D-2)\rho_{\rm r} - \biggl[(D-3) +
(D-1)\gamma^*\biggr]\rho^*\sim 0\, ,
\end{equation}
where the l.h.s of (\ref{3.10}) can be either positive or negative. So we find
that there is a range of parameters for which the CEB can be
obeyed in  some bouncing cosmologies but not in others.

In a spatially flat universe ($k=0$), the conditions for a bounce
are slightly different: $\rho_{\rm tot}=0$ and $\rho_{\rm
tot}+p_{\rm tot}<0$. At the bounce these conditions imply
violation of the Null Energy Condition (NEC). As discussed
previously, classical sources are not expected to violate the NEC, but
effective quantum sources, such as Hawking radiation, are known
to violate the NEC (see \cite{BM1,visser} for a more
comprehensive discussion of this point). In terms of $\rho_{\rm
r}$, $\rho_{\Lambda}$ and $\rho^*$ the condition for a bounce
reads
\begin{equation}
\left(1+\frac{1}{D-1}\right) \rho_{\rm r} + (1+\gamma^*)\rho^*>0.
\end{equation}
In comparison, a necessary condition that the CEB is violated can
be obtained from Eq.(\ref{cond*}),
\begin{equation}
\label{3.12}
\left(1+\frac{1}{D-1}\right) \rho_{\rm r} +
(1+\gamma^*)\rho^*\sim 0\, ,
\end{equation}
where the l.h.s of (\ref{3.12}) can be either positive or negative. 
So, again,
we find that there is a range of parameters for which the CEB can
be obeyed in  some spatially flat bouncing cosmologies but not in
others.

The CEB appears to be a more reliable criterion than energy
conditions when trying to decide whether a certain cosmology is
reasonable: taking again deSitter Universe as an example, we can
add a small amount of radiation to it, and still have a bouncing
model if $\rho_{\Lambda}$ is the dominant source, and SEC will
not be obeyed (see Eq.(\ref{condbounce})). Nevertheless, the
general discussion in this section shows that in this case the
CEB is not violated as long as radiation
temperatures remain subPlanckian, despite the presence of a
bounce. This happens, in part, because the CEB  is able to
discriminate better between dynamical and non-dynamical sources
(such as the cosmological constant), and imposes constraints that
involve the former ones only, such as Eq.~(\ref{cond*}).

\section{Conclusions}

We have reached the following conclusions by studying the
validity of the CEB for non-singular cosmologies:
 \begin{enumerate}

\item
Violation of the CEB necessarily requires either high temperatures
${\cal N} \left(\frac{T}{M_{\rm P}}\right)^{D-2} \gaq 1$, or
dynamical sources that have negative energy densities with a large
magnitude, or sources with acausal equation of state. Of course,
neither of the above is sufficient to guarantee violations of the
CEB.

\item
Classical sources of this type are suspect of being unphysical or
unstable, but each source has to be checked on a case by case
basis. In the examples we discussed in sect. II, the sources were
indeed found to be unstable or are strongly suspected to be so.

\item
Sources with large negative energy density could allow, in
principle, to increase the entropy within a given volume, while
keeping its boundary area and the total energy constant. This
would lead to violation of all known entropy bounds, and of any
entropy bound which depends in a continuous way on the total
energy or on the linear size of the system.

\item
The CEB is more discriminating than singularity theorems. In the
examples we have considered it allows non-singular cosmologies for
which singularity theorems cannot be applied, but does not allow
them if they are associated with specific dynamical problems.

\end{enumerate}

\acknowledgments
It is a pleasure to thank J. Bekenstein for
many enlightening discussions. This research was supported by
grants No. 174/00-2 (RB and SF) and No. 129/00-1 (AEM) 
of the Israel Science Foundation. SF wishes to acknowledge support from the
Kreitman Foundation.

\end{document}